# A Novel Algorithm for Compressive Sensing: Iteratively Reweighed Operator Algorithm (IROA)

Lianlin Li, and Fang Li
*Institute of Electronics, Chinese Academics of Sciences, Beijing, China*

*Abstract*- **Compressive sensing claims that the sparse signals can be reconstructed exactly from many fewer measurements than traditionally believed necessary. One of issues ensuring the successful compressive sensing is to deal with the sparsity-constraint optimization. Up to now, many excellent theories, algorithms and software have been developed, for example, the so-called greedy algorithm ant its variants, the sparse Bayesian algorithm, the convex optimization methods, and so on. The formulations for them consist of two terms, in which one is $\|\Phi u - b\|_2^2$ and the other is $\|u\|_p^p$ ( $p \leq 1$, mostly, p=1 is adopted due to good characteristic of the convex function) (NOTE: without the loss of generality, *u* itself is assumed to be sparse). It is noted that all of them specify the sparsity constraint by the second term. Different from them, the developed formulation in this paper consists of two terms where one is $\|\Phi U u - b\|_2^2$ with $U = diag(u_i^p)$ ($p \geq 1$) and the other is $\|u\|_2^2$. For each iteration the measurement matrix (linear operator) $\Phi$ is reweighed by $U$ determined by $u$ which is obtained in the previous iteration, so the proposed method is called the iteratively reweighed operator algorithm (IROA). Moreover, in order to save the computation time, another reweighed operation has been carried out; in particular, the columns of $\Phi$ corresponding to small $\{u_i\}$ have been excluded out. Theoretical analysis and numerical simulations have shown that the proposed method overcomes the published algorithms.**

*Index Terms*-**compressive sensing, the sparse signal processing, iteratively reweighed algorithm, the hard threshold algorithm, the greedy algorithm, the convex optimization, the $\ell p$-minimization**

## I. INTRODUCTION

Recent years, the well-known compressive sensing (or compressed sampling, compressive sampling, and so on) has been fast developed due to its strong potential in the field of the signal/imaging/video process, radar, remote sensing, medical imaging, wireless communication, sensor network, and so on, The basic principle is that the sparse or compressible signals can be reconstructed exactly from a suprisingly small number of linear measurements, provided that the measuemetns satify the so-called incoheherence properly. Different from the local-like signal sampling, the compressive sensing carries out the global-like signal sampling by selecting a series of suitable sampling signals; consequently, to recontructe the unknown signal deals with a non-linear optimization problem. Up to now, many efforts have been made and lots of excellent theories, algorithms, and software have been developed, for example, the so-called greedy algorithms including Matching Pursuit (MP), Compressed Sampling MP (CoSaMP), Tree-based or Model-based MP, Iteratively Hard Thresholding Algorithm (IHT), etc, the Basis Pursuit (BP) algorithm including the interior-point algorithm, SPGL1, GPSR and fixed-point continuation method, LASSO, LARS, and so on, and the sparse Bayesian algorithm and its variants.

The family of greedy algorithms has been proposed to recover the spare solution $u$ to the problem (P0) when the data satisfy certain conditions such as the coherence-based constraints or the restricted isometry property.

(P0)     $\min_{u \in \mathbb{R}^N} \|u\|_0$, subject to  $\Phi u = b$     (1.1)

As pointed by Zhang et al, these algorithms, by and large, involve solving a sequence of subspace optimization problem

$\min_u \|\Phi_T u_T - u\|_2^2$, subject to  $u_{i \notin T} = 0$     (1.2)

where *T* is an index set of the dominant components of $u$. Starting from $T = \varnothing$ and $u = 0$, the greedy algorithm iteratively adds new members to *T* (or delete undesirable members from *T*), solves (1.2) to obtain a new point $u$. Obviously, the greedy algorithms enjoy the less computational complexity and take advantage of the strong sparsity structure by specifying |*T*| in (1.2); however, they require some prior information, like the cardinality *K* of the sparse solution. In addition, these algorithms require more measurements for exact reconstruction than the BP method and sparse Bayesian algorithm. An alternative is related to the convex optimization problem, in particular, the $\ell_1$-regularized minimization problem (P1)

$$(P1) \quad \hat{u} = \min_u \left\{ \mu \|u\|_1 + \|\Phi u - b\|_2^2 \right\} \quad (1.3)$$

where $\mu > 0$. A remarkable result of Candes and Tao is that if, for example, the rows of $\Phi$ are randomly chosen, Gaussian distributed vectors, there is a constant C such that if the support of $u$ has size $K$ and $M \geq CK \log(N/K)$, then the solution to (1.3) will be exactly with overwhelming probability. In the other direction, it was shown that a nonconvex variant of basis pursuit will produce exact reconstruction with fewer measurements. Specifically, the $\ell_1$ norm is replaced with the $\ell_p (0 < p < 1)$ norm

$$\min_u \|u\|_p^p, \text{ subject to } \Phi u = b \quad (1.4)$$

That fewer measurements are required for exact reconstruction than when $p=1$ was demonstrated by numerical experiments, with random and nonrandom Fourier measurements. Obviously, the convex optimization approaches mentioned above do not require any prior information, but do not fully use the sparsity of the solution. To combine the good features of both the greedy algorithm and convex optimization approach, Zhang et al proposed the two-stage strategy, where firstly a first-order method based on shrinkage is applied to obtain an approximate solution to (1.3) and identify a working set; secondly a second-order method is applied to solve a smooth problem defined by the working set starting from the approximate solution. It should be noted that as shown by (1.1) (1.2) and (1.3) the formulations for existing approaches for compressive sensing explicitly or implicitly consist of two terms, in which one is $\|\Phi u - b\|_2^2$ and the other is $\|u\|_p^p$ with $0 \leq p \leq 1$. Obviously, all of them specify the sparsity constraint by the second term. Inspired by the iterative distorted Born approximation involved in electromagnetic inverse scattering, this paper develops a novel formulation which is

$$\min_u \left\{ \|\Phi \Lambda u - b\|_2^2 + \mu \|u\|_2^2 \right\} \quad (1.5)$$

with

$$\Lambda = diag\{u_i^p, i = 1, 2, \cdots, N\}, \quad p = 1, 2, \cdots$$

It is noted that (1) the produced measurement operator $\Phi \Lambda$ is *nonlinear* instead of linear, in other words, it is depends on $u$; (2) the proposed formulation consists of two terms where one is $\|\Phi U u - b\|_2^2$ with $U = diag(u_i^p)$ and the other is $\|u\|_2^2$. For each iteration the measurement linear operator $\Phi$ is reweighed by $U$ determined by $u$ which is obtained in the previous iteration, so the proposed method is called the iteratively reweighed operator algorithm (IROA). Theoretical analysis and numerical simulations show that our algorithm exhibits state-of-the-art performance both in terms of its speed and its ability to recover sparse signals. Even more, it can even recover signals when the measurement matrix is not satisfied by current compressive sensing theory. We would like to refer the readers to [1] for details.

## II. THE ITERATIVELY REWEIGHED OPERATOR ALGORITHM (IROA)

It should be pointed that the iteratively reweighted least squares (IRLS) approaches and their variants for solving (1.4) has been considered. These approaches are to replace the $\ell_p$ objective function in (1.4) by a weighted $\ell_2$ norm, in particular,

$$\min_u \sum_{i=1}^{N} w_i u_i^2, \text{ subject to } \Phi u = b \quad (1.6)$$

where the weights are computed from the previous iteration $u^{(n-1)}$, i.e. $w_i = \left| u_i^{(n-1)} \right|^{p-2}$.

In this paper, the new insight to the iterative reweighed approach has been investigated; in particular, the novel iterative formulation given by (1.5) is developed. From (1.5), one can draw the following conclusions:

(a) If $p$ is odd, the resulting formulation is suitable for the reconstruction of non-negative signal. Obviously, the non-negative constraint about the unknown signal has been implicitly enforced. Of course, if $p$ is even, the non-negative constraint has been deleted.

(b) The initial solution to (1.5) is specified as the vector whose entries are all *1*. Consequently, the result from the 1-iteration is the result from the $\ell_{\frac{1}{p+1}}$-regularized result.

(c) Different from the greedy algorithm where the number of column of $\Phi_T$ or the dimension of the subspace is increased with the iteration, the proposed method is on the contrary. From Fig.1, an important conclusion can be observed that with the proceeding of iteration, the value of signal where the signal actually is exist is enhanced and is impaired otherwise. In other words, the dimensional of subspace is decreased step-by-step. Inspired by this, the hard-threshold approach can be applied to fast the computation time of algorithm.
(d) Another very important conclusion is that the sparse signal can be exactly reconstructed only after several iterations, much smaller than one required by existing approaches.
(e) With step-by-step increasing $p$, one can reconstruct the sparse signal with much smaller measurements than existing approaches.

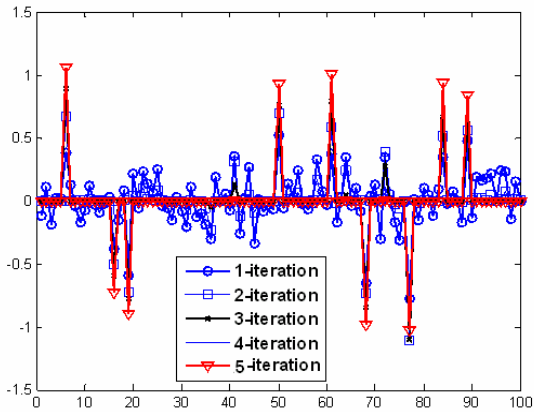

FIG.1 The iteration course of 9-sparse signal by proposed algorithm shown in Table 1. It is shown that with the proceeding of iteration, the signal is "focused" step-by-step until perfect reconstruction.

## III NUMERICAL RESULTS

For our experiments, for each of 100 trials we randomly select entries of a 50 by 200 matrix from a mean-zero Gaussian distribution, then scale the columns to have unit 2-norm. For each value of K, we randomly choose the support of $u$, then choose the sign of the components from a Gaussian distribution of mean 0 and standard derivation 1. The same $\Phi$ and $u$ would be used for each algorithm and choice of $p$. The iteration shown in Table I is run until the change in relative 2-norm from the previous iterate is less than $\varepsilon/100$ ($\varepsilon$ is prescribed parameter). Results are shown in Fig. 2 where $p=2$ is specified. We can see that the proposed IROA is able to recover signals with many more nonzero components, in comparison with the sparse Bayesian approach and iterative hard threshold algorithm.

Table I. The procedure of proposed method for Fig.1 and Fig.2

**Algorithm for Fig.1 and Fig.2**

Initialization: Choose $u^{(0)} = \mathbf{1}$ ($\mathbf{1}$ stands for the vector with element of 1)
While "*not convergence*" **do**

$$\min_u \left\{ \left\| \Phi \Lambda u^{(k+1)} - b \right\|_2^2 + \mu \left\| u^{(k+1)} \right\|_2^2 \right\}$$

Where

$$\Lambda = diag\left\{ \left[ u_i^{(k)} \right]^{p=2}, i=1,2,\cdots,N \right\}$$

Update

$$\Lambda = diag\left\{ \left[ u_i^{(k+1)} \right]^{p=2}, i=1,2,\cdots,N \right\}$$

$k \Leftarrow k+1$

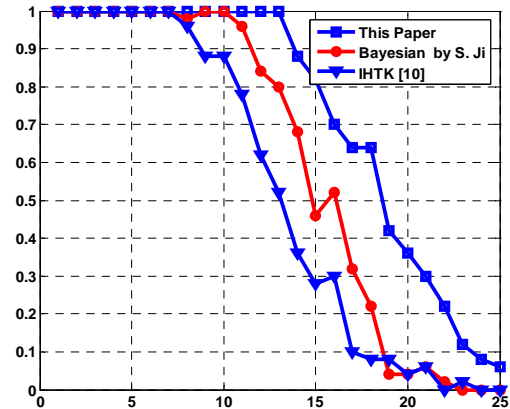

FIG.2 Plots of recovery frequency as a function of K. IROA has a much higher recovery rate than iterative hard threshold [10] and sparse Bayesian algorithm [11].

## References

[1] Lianlin Li and Fang Li, A new insight into algorithm for sparse reconstruction: the iteratively reweighed operator algorithm, in preparation.
[2] Z. Wen, W. Yin, D. Goldfarb and Y. Zhang, A fast algorithm for sparse reconstruction based on shrinkage, subspace optimization and continuation, Preprint, 2009
[3] A. M. Bruckstein, D. L. Donoho, M. Elad, From sparse solutions of systems of equations to sparse modeling of signals and images, SIAM Review, 51(1), 34-81, 2009
[4] R. Chartrand and W. Yin, Iteratively reweighed algorithms for compressive sensing, Preprint, 2008
[5] R. Chartrand, "Exact reconstruction of sparse signals via nonconvex minimization," IEEE Signal Process. Lett., vol. 14, pp. 707–710, 2007.
[6] B. D. Rao and K. Kreutz-Delgado, "An affine scaling methodology for best basis selection," IEEE Trans. Signal Process., vol. 47, pp. 187–200, 1999.
[7] E. J. Cand`es, M. B. Wakin, and S. P. Boyd, "Enhancing sparsity by reweighted `1 minimization." Preprint.

[8] D. L. Donoho, "Compressed sensing," IEEE Trans. Inf. Theory, vol. 52, pp. 1289–1306, 2006.
[9] E. J. Cand`es, J. Romberg, and T. Tao, "Robust uncertainty principles: Exact signal reconstruction from highly incomplete frequency information," IEEE Trans. Inf. Theory, vol. 52, 2006.
[10] T. Blumensath and M. E. Davies, Normalised iterative hard thresholding: guaranteed stability and performance, Preprint, 2009
[11] S. Ji, Y. Xue and L. Carin, Bayesian Compressive Sensing, IEEE Trans. on Signal Processing vol. 56 (6): 2346-2356, 2008.
[12] R. A. DeVore, "Deterministic construction of compressed sensing matrices." Preprint.
[13] J. A. Tropp and A. C. Gilbert, "Signal recovery from partial information via orthogonal matching pursuit." Preprint.
[14] S. S. Chen, D. L. Donoho, and M. A. Saunders, "Atomic decomposition by basis pursuit," SIAM J. Sci. Comput., vol. 20, pp. 33–61, 1998.
[15] E. Cand`es and T. Tao, "Near optimal signal recovery from random projections: universal encoding strategies?," IEEE Trans. Inf. Theory, vol. 52, pp. 5406–5425, 2006.
[16] D. L. Donoho and J. Tanner, "Thresholds for the recovery of sparse solutions via L1 minimization," in 40th Annual Conference on Information Sciences and Systems, pp. 202–206, 2006.
[17] C. La and M. N. Do, "Signal reconstruction using sparse tree representations," in Wavelets XI, vol. 5914, SPIE, 2005.
[18] R. Chartrand and V. Staneva, "Restricted isometry properties and nonconvex compressive sensing." Preprint.
[19] I. F. Gorodnitsky and B. D. Rao, "Sparse signal reconstruction from limited data using FOCUSS: a reweighted minimum norm algorithm," IEEE Trans. Signal Process., vol. 45, pp. 600–616, 1997.
[20] G. W. Stewart, "On scaled projections and pseudoinverses," Linear Algebra Appl., vol. 112, pp. 189–194, 1989.